\documentclass[%
 aps,prl,
rsi,%
 amsmath,amssymb,
 reprint,%
superscriptaddress
]{revtex4-1}

\usepackage{graphicx}
\usepackage{dcolumn}
\usepackage{bm}
\usepackage{color}

\begin{document}

\title{Novel catchbond mediated oscillations in motor-filament complexes}

\author{Sougata Guha}
\email{sougataguha@iitb.ac.in}
\affiliation{Department of Physics, Indian Institute of Technology Bombay, Mumbai, India}
\affiliation{Department of Physics, Savitribai Phule Pune University, Pune, India}
\author{Mithun K. Mitra}
\email{mithun@phy.iitb.ac.in}
\affiliation{Department of Physics, Indian Institute of Technology Bombay, Mumbai, India}
\author{Ignacio Pagonabarraga}
\email{ipagonabarraga@ub.edu}
\affiliation{ CECAM, Centre Europ\'een de Calcul Atomique et Mol\'eculaire, \'Ecole Polytechnique F\'ed\'erale de Lasuanne (EPFL), Batochime, Avenue Forel 2, 1015 Lausanne, Switzerland}
\affiliation{ Departament de F\'{\i}sica de la Mat\`eria Condensada, Universitat de Barcelona, Mart\'{\i} i Franqu\`es 1, E08028 Barcelona, Spain}
\affiliation{UBICS University of Barcelona Institute of Complex Systems,  Mart\'{\i} i Franqu\`es 1, E08028 Barcelona, Spain}
\author{Sudipto Muhuri}
\email{sudipto@physics.unipune.ac.in}
\affiliation{Department of Physics, Savitribai Phule Pune University, Pune, India}

\date{\today}

\begin{abstract}
Generation of mechanical oscillation is ubiquitous to wide variety of intracellular processes. We show that catchbonding behaviour of motor proteins provides a generic mechanism of generating spontaneous oscillations in motor-cytoskeletal filament complexes. We obtain the phase diagram to characterize how this novel catch bond mediated mechanism can give rise to bistability and sustained limit cycle oscillations and results in very distinctive stability behaviour, including bistable and non-linearly stabilised in motor-microtubule complexes in biologically relevant regimes. Hitherto, it was thought that the primary functional role of the biological catchbond was to improve surface adhesion of bacteria and cell when subjected to external forces or flow field. Instead our theoretical study shows that the imprint of this catch bond mediated physical mechanism would have ramifications for whole gamut of intracellular processes ranging from oscillations in mitotic spindle oscillations to activity in muscle fibres.

\end{abstract}

\maketitle
 Occurrence of spontaneous oscillations due to molecular motor activity is germane to a variety of intracellular processes - ranging from mitotic cell division process \cite{spindle1,karsten1,raja,paolo} to oscillations in muscle fiber~\cite{karsten2,sasaki}. While for muscle fibers, mechanical oscillations arise due to force dependent detachment of myosin motors to actin filaments~\cite{karsten2}, spindle oscillations are generated during cell division  due to the interplay of the microtubule (MT) filament elasticity with stochastic (un)binding of cortical dynein motors to(off) these filaments~\cite{karsten1}. Unravelling the underlying physical mechanism by which oscillations are generated in such motor-biofilament complexes is essential to understanding the spatio-temporal organization of the cell and their stability characteristics. 

The unbinding characteristics of individual motors constitutes a crucial determinant of the collective properties of motor-filament complexes \cite{joanny1, manoj, chou, sougata-epl, debasoft,  lipo-bd, xu,igna-pre}. Dynein motors - which walk along  MTs, and myosin-II - which walk along actin, exhibit  catchbonding \cite{kunwar,guo}, where the motor unbinding rate decreases when subject to increasing load forces. This non-trivial force dependence in the unbinding rates leads to novel collective behavior in multiple motor-filament complexes, in contrast to slip-bonded motors such as kinesin \cite{pr-res, anil}. A striking consequence is known as the paradox of codependence - where inhibition of one species of motor in bidirectional intra-cellular transport can lead to an overall decline in the motility of the cellular cargo \cite{pr-res,hancock}. The implications of this ubiquitous feature of catchbonding on the functional behavior has not been studied in context of the mechanical behaviour of these complexes.

In this letter, we study the implications of the catchbonded behavior of motors on  the stability of motor-filament complexes. We find that the catchbonded nature of the motor unbinding process leads to a generic mechanism of generation of spontaneous oscillations  and results in very distinctive stability behaviour, including bistable and non-linearly stabilised  motor-filament complexes. We argue that this feature has important ramifications for the stability behaviour and oscillations in mitotic spindles, where such motor-microtubules complexes are the primary constituents of the spindle structure. 

\begin{figure}[t!]
	\centering
	\includegraphics[width=\linewidth]{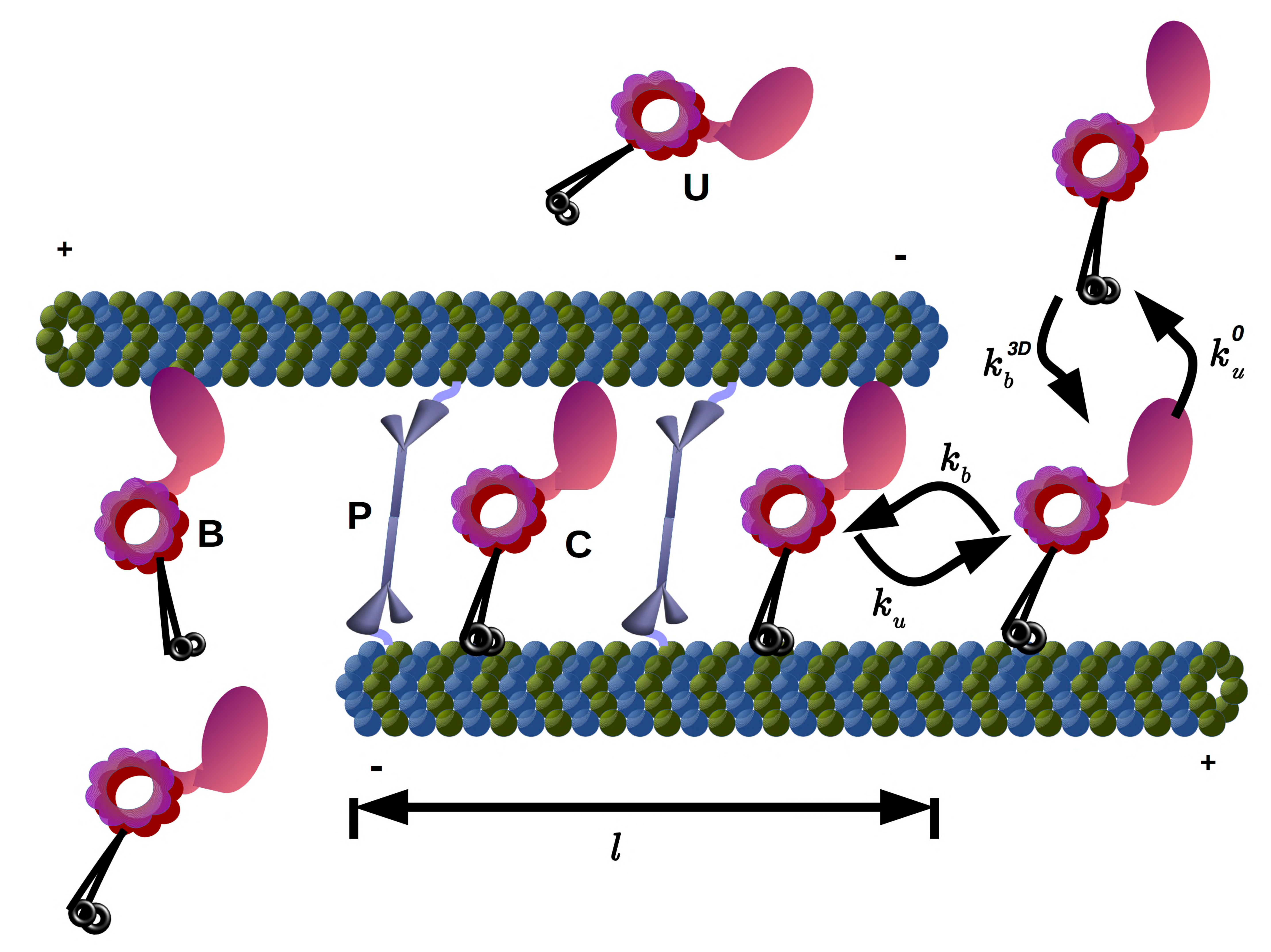}
	\caption{Schematic diagram of the antiparallel MT-motor complex in presence of passive crosslinkers (P). Unbound motors (U) attach to any  of the MTs with rate $k_b^{3D}$,  bound motors (B) crosslink at rate $k_b$ whereas they detach from the MT at a rate $k_u^0$. Crosslinked motors (C) become  bound motors with detachment rate $k_u$ under a load force $F_p$. }
	\label{fig:Fig1}
\end{figure}

In order to shed light on the generic mechanism of generation of spontaneous oscillations and analyze the stability behaviour in such motor-microtubule complexes, we adopt a minimalist approach to describe the stability  of a specific system comprising of a pair of overlapping MTs, dynein motors and confined passive crosslinkers (Fig.~\ref{fig:Fig1}). For this minimal arrangement, dynein motors in the overlap region can crosslink the MTs and generate sliding forces which  tend to decrease the overlap length ($l$), while the confined passive proteins (P)  generate an entropic force which tends to increase the overlap length~\cite{lansky}. The force exerted by $N_p$  passive proteins confined to the overlap region reads  $F_p = N_{p}\epsilon/l$, where $\epsilon$ is the  passive proteins binding energy. We distinguish between two population of motors: i) $n_c$ crosslinked motors - which are bound to both the filaments and hence cause mutual sliding of the filaments, while experiencing a load force $F_p$, and ii) $n_b$ bound motors - which are bound to only one of the MT filaments, and hence exerts no sliding force nor feel the effect of the force due to passive proteins. The dynamic equation for $l$ can be expressed as,
 \begin{eqnarray}
\frac{dl}{dt} &=& -2v_{0} ~\Theta~ (n_c f_s - F_p) \left( 1 - \frac{F_{p}}{n_{c}f_{s}} \right) + \frac{F_p}{\Gamma}
\end{eqnarray}
where $f_s$ is the stall force for single dynein motor, $v_0$ the single dynein velocity in the absence of load force, and  $\Gamma$ is the friction constant. For simplicity,  we have assumed a linear force-velocity relation for the crosslinked dynein motors~\cite{lipo-uni}, zero backward velocity of these crosslinked motors in superstall conditions~\cite{lipo-bd}, and that the load force $F_p$ due to passive proteins is shared equally by the crosslinked dynein motors~\cite{lipo-uni,lipo-bd}. The Heaviside, $\Theta$ function ensures that motor sliding  contributes to the change in $l$ only below the stall condition for forces on the crosslinked motors, whereas above stall the dynamics of the overlap length is governed only by the entropic forces exerted by the confined proteins. 

Motor (un)binding  kinetics in the overlap region is expressed in terms of  rate equations. The crosslinked dynein motors are subject to $F_p$ and they can exhibit catchbond behaviour beyond the threshold force, $f_m$~\cite{kunwar}.  Incorporating  dynein catchbonding   in a phenomenological threshold force bond deformation model~\cite{anil,pr-res}, the crosslinked dynein unbinding rate reads,
 \begin{equation}
    k_{u} = n_{c} k_{u}^{0} \exp [-E_d(F_p) + F_p/(n_{c} f_d) ~]\nonumber
  \end{equation}
where $f_d$ denotes the characteristic dynein detachment force   in the slip region ($F_p < n_c f_{m}$), and the deformation energy $E_d$ is activated when $F_p > n_c f_{m}$, reads~\cite{anil},
\begin{equation}
    E_d(F_p) = \Theta (F_{p}  - n_{c} f_m)~ \alpha \left[1 - \exp\left(-\frac{F_p/n_{c} - f_m}{f_0}\right)\right]\nonumber
\end{equation}
\begin{figure}[t]
    \centering
    \includegraphics[width=\linewidth]{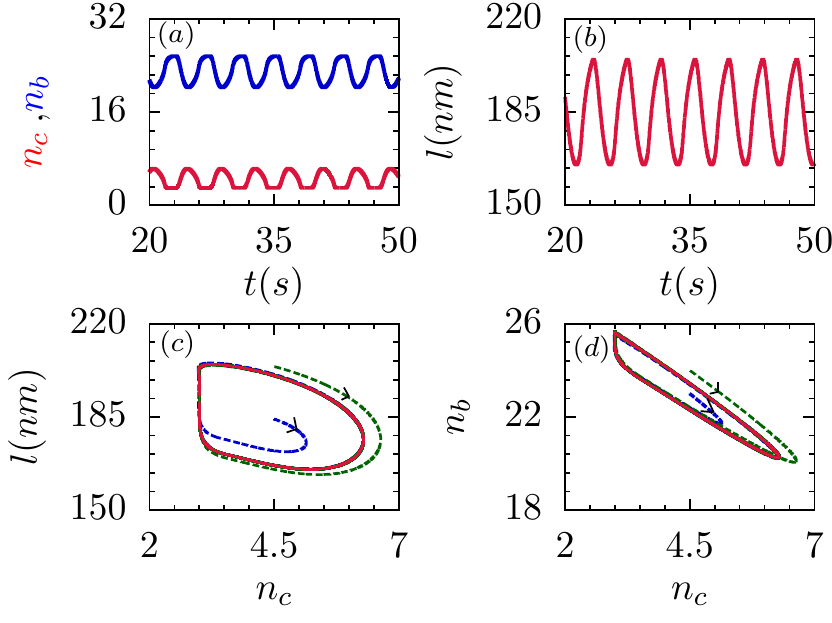}
    \caption{Limit cycle oscillations for  (a) $n_c$ (red curve) and $n_b$ (blue curve)  and  (b) $l$ as a function of time. The figures show that these quantities oscillate between $n_c \sim 3-6$, $n_b \sim 20-26$  and $l \sim 164-205 $ nm.  (c) and (d) depict the variation of $l$ and $n_b$ with $n_c$, respectively (red curves).  The  blue and green dashed lines depict two sample trajectories with different initial conditions eventually falling onto the limit cycle.  All the curves are obtained for $\tilde{f}_s=0.76$, $\Delta_n=6$, which corresponds to a point denoted by `$\blacklozenge$' in the limit cycle region in Fig.~\ref{fig:Fig2}(d).} 
    \label{fig:Fig3}
\end{figure}   

where, $\alpha$ measures the  catchbond strength, while $f_0$ denotes the force scale associated with the catchbond deformation energy. The dynamic equation for $n_c$ is,
\begin{equation}
\frac{dn_c}{dt} = k_{b}n_{b} - k_{u}^{0} n_{c}\exp (\eta) 
\end{equation}
with,
\begin{equation}
\eta =  \frac{F_{p}}{n_{c} f_{d}} -\Theta \left( \frac{F_p}{n_c} - f_m \right) \alpha \left[ 1 - \exp \left( -\frac{F_p/n_{c} - f_m}{f_0} \right)  \right] \nonumber
\end{equation}
Here, $k_{b}$ and $k_{u}^{o}$ are the rates with which the bound motors are converted to crosslinked state and vice-versa, in the absence of external load force. Specifically, bound motors are lost due to conversion to crosslinked motors (rate $k_b$), and due to unbinding from the MT filament (rate $k_u^0$), while the gain terms are due to conversion from crosslinked motors (rate $k_u^0$), binding of free motors from the bulk onto the overlap region of the filament (rate $k_b^{3D}$), and due to the incoming flux ($J$) of bound motors  from the two ends of the overlap region. The corresponding dynamic equation for $n_{b}$ is, 
\begin{eqnarray}
\frac{dn_b}{dt} = k_{u}^{o}n_{c}\exp(\eta) - (k_{u}^{o} + k_{b})n_{b} + k_{b}^{3D}\rho_{3d}l + 2J 
\end{eqnarray}
where, $\rho_{3d}$ is the bath  dynein motor linear density.
In the limit of $l$ being much smaller than the MT length, the incoming flux from a single end is $J = \frac{ k_{b}^{3d}\rho_{3d}}{k_{u}^{0}}\left[v_0 + \frac{dl}{dt}\right]$ \cite{ignaref, igna-epl}.
The equations are analysed using the rescaled dimensionless variables 
$\tilde{l} = \tilde{f_s}l/l_{p}~$, $~l_p = v_0/k_{u}^{0}~$,$~\tilde{f_s} = \frac{bf_s}{k_{B}T}~$,$~\tau = t k_{u}^{0}~$,$~l_e = \frac{b \epsilon}{k_{B}T}~$,$~\zeta = l_e/l_p~$, $~\tilde{\Gamma} = \frac{2bv_{0}\Gamma}{k_{B}T}~$,$~\tilde{f_0} = \frac{b f_0}{k_{B}T}$,$~\tilde{f_d} = \frac{bf_d}{k_{B}T}~$,$~\tilde{f}_m = \frac{bf_m}{k_{B}T}~$,$~\Delta_n = \frac{k_{b}^{3D} \rho_{3D} v_0}{k_{u}^{0}~k_{u}^{0}}$,$~\gamma~=~k_b/k_u^0$. The corresponding dimensionless equations are provided in the Supplementary Material.


\begin{figure*}[t]
    \centering
    \includegraphics[width=\linewidth]{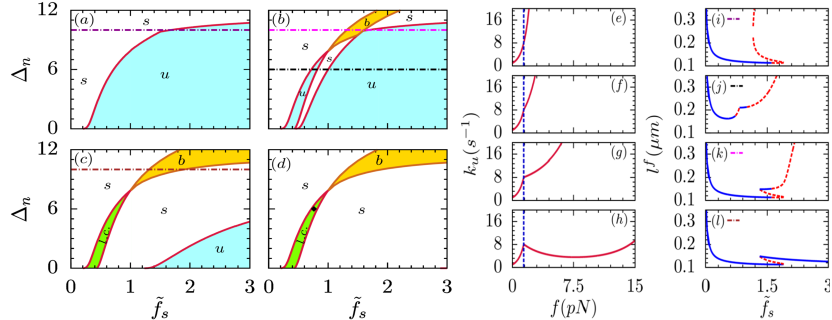}
    \caption{ Stability diagram of a MT-motor complex as a function of  $\tilde{f_0}$. (a) $\tilde{f_0}=309.52~ (f_o \approx 1000$ pN), (b) $\tilde{f_0}=24.76 ~(f_o \approx 80$ pN), (c) $\tilde{f_0}= 15.47 ~(f_o \approx 50$ pN) and (d) $\tilde{f_0}=11.98~ (f_o \approx 38.7$ pN). All other parameters are $N_p=100, k_b=k_u^0=1~/s~ (\therefore \gamma=1), v_0=100$ nm/s, $\tilde{\Gamma}=2.6, \tilde{f_d}=0.2~ (f_d \approx 0.67$ pN), $\alpha=68, \tilde{f_m}=0.43 ~(f_m \approx 1.4$ pN), $b=1.3$ nm, $\epsilon = 2k_BT$.The red solid line depicts the boundary between linearly stable (white) and unstable (cyan) regions.  Green areas  indicate regions where limit cycles can be sustained and yellow shaded areas signal regions where the complex displays bistable behaviour. Panels (e-h) depict the unbinding rate of a single dynein motor under load force, $f$, for the  $f_0$ values of panels a-d, while  blue dashed line gives the reference curve when $\tilde{f_s}=\tilde{f_m}$ 
    Panels (i-l) show the bifurcations diagrams as a function of $\tilde{f_s}$ in different regions of the phase plane, as indicated by the dashed lines in panels (a-c). The solid blue lines indicate a stable branch, while the dashed red lines indicate unstable solutions.}
    \label{fig:Fig2}
\end{figure*}  

The main striking feature, when filaments are cross-linked by catchbonded motors, is the generation of sustained limit cycle oscillations, as shown in Fig.~\ref{fig:Fig3}.
These oscillations emerge exclusively from   catch-bonding, and result from the non-linear stabilization of previously unstable morphologies. The corresponding sustained oscillations of $n_c$ and $n_b$ are displayed in Fig.~\ref{fig:Fig3}(a) and those of  $l$ in Fig.~\ref{fig:Fig3}(b), while  Fig.~\ref{fig:Fig3}(c) and (d) display the limit cycle behaviour in the $n_c - l$ and the $n_c - n_b$ planes, respectively. Qualitatively, in  the linearly unstable region, the motor force imbalance increases $l$, while $n_c$ decreases due to the increased propensity of the crosslinked motors to detach, corresponding to slip behaviour. As $n_c$ further decreases,  motor loading forces increase leading to  motor catchbonding, prolonging  their attachment, favouring  reattachment, and eventually counterbalancing the entropic forces from passive confined proteins. This arrests the  increase in $l$ and leads to the overall force due to crosslinking motors overpowering the force due to passive proteins, leading to a decrease of the overlap length. Effectively, dynein catchbonding  results in a negative feedback loop which leads to limit cycle oscillations. Catchbond-mediated oscillations offers a hitherto unappreciated mechanism through which mechanical oscillations can be generated in motor-filament complexes.

To comprehensively  quantify  the effect of catchbonding, we analyze the stability of  motor-filament complexes for biologically relevant regimes by using available experimental data for single motors and passive proteins. Since the single dynein motor velocity, $v_o = 0.1~\mu$m$s^{-1}$~\cite{vale-cell}, and bare unbinding rate  $k_u^0 = 1~s^{-1}$~\cite{vale-cell},  then  $l_p = 0.1 ~\mu m$. 
For a characteristic thermal energy, $k_B$T$=4.2$ pN-nm and motor unbinding length scale, $b = 1.3$ nm~\cite{schnitzer}, one gets $l_e = 2.6$~nm. The stall force for cytoplasmic dynein, which can be modulated by using dynactin and BICD2N complexes~\cite{belyy}, has reported values $f_s= 1.25$ pN~\cite{roop2005}, leading to  $\tilde f_s \sim 0.39$, while for yeast dyenin, $f_s= 7$pN~\cite{higuchi2006}, corresponding to $\tilde f_s \sim 2.17$. Reported binding rates, $k_b = 1~s^{-1}$\cite{leduc} and $k_b^{3D}\rho_{3D}=60~\mu$m$^{-1}s^{-1}$, lead to $\Delta_n = 6$, which may vary at least over a range $0.5-10$ on varying physiological conditions, while estimated  passive crosslinker binding energies, $\epsilon=2k_B$T~\cite{sougata-epl}. Quantitative estimates of the friction coefficient of passive crosslinkers propose $\tilde{\Gamma}~=~2.6$ for a few hundred passive proteins~\cite{lansky}. Finally, the observed unbinding rates of single dynein for different load forces can be replicated with catchbond strengths $\tilde{f_0} = 11.98$, and $\alpha= 68$, $\tilde{f_d} = 0.2$~\cite{anil,kunwar}. The full list of model parameters is listed in Table I (Supplementary Material).


%

Fig.~\ref{fig:Fig2}  displays  the stability diagram of motor-MT complexes  as  $f_0$ is varied for a fixed $\alpha$, in the $\Delta_{n}- \tilde f_s$ plane, where $\Delta_{n}$ is a tunable biological parameter associated with the propensity of the motor to bind to the MT filament (Fig.~\ref{fig:Fig2}), when the catchbonding force threshold differs from the motor stall force, i.e. $f_m \neq f_s$. In the absence of catchbonding, ($f_0 \rightarrow \infty$) the phase diagram  has just two morphologies corresponding to  a  linearly stable and  unstable overlapping MTs. This is shown in Fig. \ref{fig:Fig2}(a). The corresponding bifurcation diagram  is shown in Fig.~\ref{fig:Fig2}(i) for $\Delta_n = 10$, and the force-dependent unbinding rate in this regime is shown in Fig.~\ref{fig:Fig2}(e).

The effect of the catchbond strength can be assessed  varying  $f_0$. For  weak catchbonding ($\tilde{f_0} = 24.76$),  the unbinding rate  increases with opposing load forces, although it exhibits  a kink at $f = f_m$, for which the rate of increase of unbinding rate is slower owing to  catchbonding (see Fig.~\ref{fig:Fig2}(f)), as  shown in Fig.~\ref{fig:Fig2}(b). For   small  $\Delta_{n}(\Delta_{n} \lesssim 8$), the complex first becomes unstable, as in the non-catchbonded regime. On increasing $\tilde{f_s}$ further, the system shows a re-entrant transition  where the complex is stable, before finally becoming unstable at higher $\tilde{f_s}$.  As Fig.~\ref{fig:Fig2}(j) displays the corresponding bifurcation diagram  for $\Delta_n=6$, the  system always has one single fixed point. As $\tilde{f_s}$ is increased, this fixed point changes stability from stable to unstable to stable to unstable through a series of bifurcations (Hopf, saddle and Hopf). This re-entrant behaviour can be  qualitatively understood because at  intermediate $\tilde{f_s}$, the load force on individual motors is high enough for them to be catchbonded and thus the unbinding rate is relatively small compared to pure slip. This leads to a higher number of motors attaching to the filament than in absence of catchbonding, which in turn implies that the the sliding forces exerted by the motors counterbalance  the passive crosslinker force,  resulting in  complex stabilization. At higher values of $\tilde{f_s}$, the unbinding rate is sufficiently high, and  the remaining motors can no longer stabilize the MT-motor complex. For larger $\Delta_{n}$ (  $\Delta_{n} \gtrsim 8$), the complex displays  bistability for a range of $\tilde{f_s}$. A representative bifurcation diagram is shown in Fig.~\ref{fig:Fig2}(k) at $\Delta_n = 10$. For small $\tilde{f_s}$, the complex displays a single stable fixed point. Beyond a critical $\tilde{f_s}$, the system undergoes a saddle-node bifurcation leading to the emergence of a new stable fixed point, in addition to an unstable fixed point. This corresponds to the region of bistable behaviour. On increasing  $\tilde{f_s}$ further, the new  fixed point destabilizes via a Hopf bifurcation, leading to a single stable steady state. At even larger  $\tilde{f_s}$, this stable fixed point disappears via a reverse saddle-node bifurcation and the system become unstable. This bistable behaviour arises due to the catchbonded nature of the unbinding characteristics of dynein motors.

On decreasing $\tilde f_0$ further, not only does the linearly stable region become larger, but  the region of linearly unstable configurations ( for intermediate values of $\tilde f_s$) are stabilized by a non-linear mechanism leading to limit-cycle oscillations. This non-linear stabilization arises purely due to dynein catchbonding. The bistable regions also grow with decreasing $\tilde{f_0}$, as  shown in Fig.~\ref
{fig:Fig2}(c) for $\tilde{f_0}=15.47$. For even stronger catchbonding, where the unbinding rate decreases sharply beyond $f_m$ (see Fig.~\ref{fig:Fig2}(d)), the nature of the phase diagram remains similar, except that  the region of stable overlaps grows even bigger, as  shown in Fig.~\ref{fig:Fig2}(d) for $\tilde{f_0}=11.98$~\cite{anil}. These limit cycle oscillations are robust even under fluctuations of the order of the underlying energy scales in the system (see Supplementary Material).

\begin{figure}[t]
    \centering
    \includegraphics[width=\linewidth]{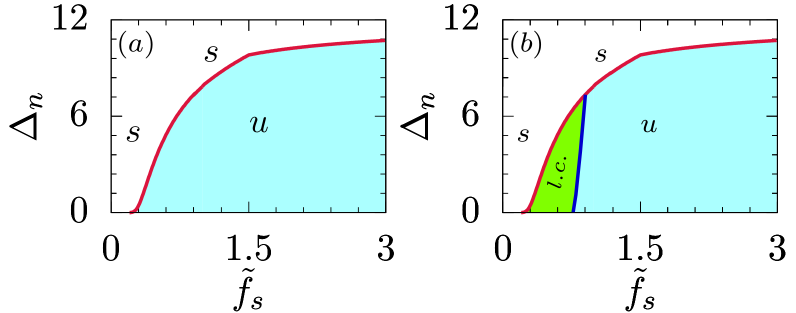}
    \caption{Stability diagram when $\tilde{f}_s=\tilde{f}_m$,  (a) in the absence of catchbond ($\tilde{f_0}=309.52$), and (b) presence of catchbond, ($\tilde{f_0} = 11.98$). All other parameters are same as Fig.~\ref{fig:Fig2}. The red solid curve corresponds to  the boundary between linearly stable (white) and linearly unstable regions (cyan). The figure shows that limit cycles (green) develop only due to catchbond.}
    \label{fig:Fig6}
\end{figure}   

Experimental studies  suggest that for dynein, catchbond sets in  around the motor stall force, $f_m\simeq f_s = 1.25$ pN~\cite{kunwar,pr-res}. Fig.~\ref{fig:Fig6} displays the phase diagram in the $\Delta_n - \tilde{f_s}$ plane, when $f_s=f_m$. In the absence of catchbonding, Fig.~\ref{fig:Fig6}(a), a  stable-to-unstable transition is recovered, cf.  Fig.~\ref{fig:Fig2}(a). In the presence of catchbonding, e.g; decreasing $f_o$,  limit-cycle oscillations appear in the phase diagram, see Fig.~\ref{fig:Fig2}(b), where  unstable overlaps are non-linearly stabilized. The  region with  oscillatory behaviour increases with enhanced catchbonding, i.e. decreasing $f_0$.
 
For experimentally relevant parameters, the limit cycle oscillations have a typical time period  $\sim 1-20s$, amplitude $\sim 0.05 - 0.2 ~\mu m$, and characteristic   overlap length in the range of $0.1 - 1 ~\mu m$ (See Supplementary Material). These estimates  lie  within the observable temporal and spatial scale of cellular processes and points to their biological relevance in context of motor-microtubule complexes in particular and more generally in context of mitotic spindles.

In summary, in this Letter, we  investigated the functional consequences of   molecular motor's catchbonding on the stability  of motor-biofilament complexes. We find that  for a pair of overlapping antiparallel biofilaments subject to sliding forces by  motors and entropic forces by confined passive crosslinkers, the catchbond  nature of motor unbinding from the biofilaments  manifests  as a generic intrinsic mechanism that generates and stabilizes  spontaneous oscillations, and additionally also promotes  bistability in biologically relevant regimes. 

Recent experiments report that kinesin motors exhibit  catchbonding under horizontal load forces~\cite{howard-catchbond}; hence,  it would be interesting to analyze whether catchbond-driven oscillations are present for such  kinesin-MT complexes. Controlled experiments on immobilized MTs on glass surfaces~\cite{lansky} offer  a  promising experimental setup to  verify the described collective  implications of   catchbonding on motor-filament complexes.

The mechanism of nonlinear oscillations we have described   is distinct from previously reported  oscillation mechanisms for MT-motor complexes that arise from the coupling of  motor proteins in the cell cortex with the overlapping MTs~\cite{debasoft, karsten}, and shown to be relevant for understanding mitotic oscillations in spindles~\cite{debasoft, karsten1}. The  range of oscillation frequency predicted (0.1-1 Hz) lie in the same range of previously reported mechanisms  and in the experimentally observed  oscillation frequency range in the mitotic spindle during the metaphase of cell division~\cite{PecreauxCurrBiol2006}. Since the mitotic spindle, in the metaphase, is composed of  overlapping MTs that interact with  cortical motor proteins,  and are  subject to  sliding forces of crosslinking motors e.g; dynein and Eg5 kinesin, as well as  to kinetochores and chromosomes~\cite{paolo,raja}, clarifying whether these distinct oscillation mechanisms can result in resonances,  with their potential implications for spindle  stability, remains an open challenge.









\noindent
{\em Acknowledgements.} Financial support is acknowledged by MKM for Ramanujan Fellowship (13DST052), DST and  IITB (14IRCCSG009); SM and MKM for SERB project No. EMR /2017/001335]. I.P. acknowledges the support from MINECO/FEDER UE (Grant No. PGC2018-098373-B-100), DURSI (Grant No. 2017 SGR 884), and SNF Project no. 200021-175719.

SG analyzed the model and data, performed the numerics and wrote the manuscript. MKM analyzed the model and data, and wrote the manuscript. IP and SM designed the study, proposed the model, analyzed the model and data, and wrote the manuscript.

\appendix


\begin{thebibliography}{10}

\bibitem{spindle1} D. G. Albertson, Developmental Biology {\bf 101}, 61 (1984).  

\bibitem{karsten1} S. W. Grill, K. Kruse, and F. J\"ulicher, Physical Review
Letters {\bf 94}, 108104 (2005).

\bibitem{raja} N. P. Ferenz, R. Paul, C. Fagerstrom, A. Mogilner, and
P. Wadsworth, Current Biology {\bf 19}, 1833 (2009).

\bibitem{paolo} P. Malgaretti and S. Muhuri, EPL (Europhysics Letters) {\bf 115}, 28001 (2016).

\bibitem{karsten2} S. G\"unther and K. Kruse, New Journal of Physics {\bf 9}, 417 (2007).

\bibitem{sasaki} D. Sasaki, H. Fujita, N. Fukuda, S. Kurihara, and S. Ishiwata, Journal of Muscle Research and Cell Motility {\bf 26}, 93 (2005).

\bibitem{joanny1} T. Gu{\'e}rin, J. Prost, P. Martin, and J.-F. Joanny, Current Opinion in Cell Biology {\bf 22}, 14 (2010).

\bibitem{manoj} D. Bhat and M. Gopalakrishnan, Physical Biology {\bf 9}, 046003 (2012).

\bibitem{chou} F. Posta, M. R. D'Orsogna, and T. Chou, Physical Chemistry Chemical Physics {\bf 11}, 4851 (2009).

\bibitem{sougata-epl} S. Guha, S. Ghosh, I. Pagonabarraga, and S. Muhuri, EPL (Europhysics Letters) {\bf 124}, 58003 (2019).

\bibitem{debasoft} S. Ghosh, V. N. S. Pradeep, S. Muhuri, I. Pagonabarraga, and D. Chaudhuri, Soft Matter {\bf 13}, 7129 (2017).

\bibitem{lipo-bd} M. J. I. M{\"u}ller, S. Klumpp, and R. Lipowsky, Proceedings of the National Academy of Sciences {\bf 105}, 4609 (2008).

\bibitem{xu} D. Ando, M. K. Mattson, J. Xu, and A. Gopinathan, Scientific Reports {\bf 4}, 1 (2014).

\bibitem{igna-pre} S. Muhuri and I. Pagonabarraga, Physical Review E {\bf 82}, 021925 (2010).

\bibitem{kunwar} A. Kunwar, S. K. Tripathy, J. Xu, M. K. Mattson,
P. Anand, R. Sigua, M. Vershinin, R. J. McKenney, C. Y.
Clare, A. Mogilner, and S. P. Gross, Proceedings of the National Academy of Sciences {\bf 108}, 18960 (2011).

\bibitem{guo} B. Guo and W. H. Guilford, Proceedings of the National Academy of Sciences {\bf 103}, 9844 (2006).

\bibitem{pr-res} P. Puri, N. Gupta, S. Chandel, S. Naskar, A. Nair,
A. Chaudhuri, M. K. Mitra, and S. Muhuri, Physical Review Research {\bf 1}, 023019 (2019).

\bibitem{anil} A. Nair, S. Chandel, M. K. Mitra, S. Muhuri, and A. Chaudhuri, Physical Review E {\bf 94}, 032403 (2016).

\bibitem{hancock} W. O. Hancock, Nature Reviews Molecular Cell Biology {\bf 15}, 615 (2014).

\bibitem{lansky} Z. Lansky, M. Braun, A. L{\"u}decke, M. Schlierf, P. R. ten
Wolde, M. E. Janson, and S. Diez, Cell {\bf 160}, 1159 (2015).

\bibitem{lipo-uni} S. Klumpp and R. Lipowsky, Proceedings of the National Academy of Sciences {\bf 102}, 17284 (2005).

\bibitem{ignaref} O. Camp{\`a}s, J. Casademunt, and I. Pagonabarraga, EPL
(Europhysics Letters) {\bf 81}, 48003 (2008).

\bibitem{igna-epl} S. Muhuri, I. Pagonabarraga, and J. Casademunt, EPL (Europhysics Letters) {\bf 98}, 68005 (2012).

\bibitem{vale-cell} S. L. Reck-Peterson, A. Yildiz, A. P. Carter, A. Gennerich, N. Zhang, and R. D. Vale, Cell {\bf 126}, 335 (2006).

\bibitem{schnitzer} M. J. Schnitzer, K. Visscher, and S. M. Block, Nature
Cell Biology {\bf 2}, 718 (2000).

\bibitem{belyy} V. Belyy, M. A. Schlager, H. Foster, A. E. Reimer, A. P. Carter, and A. Yildiz, Nature Cell Biology {\bf 18}, 1018 (2016).

\bibitem{roop2005} R. Mallik, D. Petrov, S. Lex, S. King, and S. Gross, Current Biology {\bf 15}, 2075 (2005).

\bibitem{higuchi2006} S. Toba, T. M. Watanabe, L. Yamaguchi-Okimoto, Y. Y.
Toyoshima, and H. Higuchi, Proceedings of the National Academy of Sciences {\bf 103}, 5741 (2006).

\bibitem{leduc} C. Leduc, O. Camp{\`a}s, K. B. Zeldovich, A. Roux, P. Jolimaitre, L. Bourel-Bonnet, B. Goud, J.-F. Joanny, P. Bassereau, and J. Prost, Proceedings of the National Academy of Sciences {\bf 101}, 17096 (2004).

\bibitem{howard-catchbond} H. Khataee and J. Howard, Physical Review Letters {\bf 122}, 188101 (2019).

\bibitem{karsten} D. Johann, D. Goswami, and K. Kruse, Physical Review Letters {\bf 115}, 118103 (2015).

\bibitem{PecreauxCurrBiol2006} J. Pecreaux, J. -C. R{\"o}per, K. Kruse, F. J{\"u}licher, A. A. Hyman, S. W. Grill, and J. Howard, Current Biology {\bf 16}, 2111 (2006).

\end{thebibliography}
\end{document}